\documentclass[fleqn,10pt]{wlscirep}
\usepackage[utf8]{inputenc}
\usepackage[T1]{fontenc}
\graphicspath{ {figures/} }
\usepackage{diagbox}
\usepackage{makecell}
\usepackage{comment}
\usepackage[skip=6pt plus 2pt]{parskip}
\usepackage{float}
\usepackage{xr}
\usepackage{caption} 
\usepackage{titlesec}

\titlespacing*{\section}
{0pt}{5ex plus 1ex minus .2ex}{4ex plus .3ex}
\titlespacing*{\subsection}
{0pt}{4.5ex plus 1ex minus .2ex}{3ex plus .2ex}
\titlespacing*{\subsubsection}
{0pt}{4ex plus 1ex minus .2ex}{2.5ex plus .2ex}

\def\*#1{\mathbf{#1}}
\makeatletter

\title{Two pathways to resolve relational inconsistencies}

\author[1,*]{Tomer Barak}
\author[1,2]{Yonatan Loewenstein}
\affil[1]{The Edmond and Lily Safra Center for Brain Sciences, The Hebrew University, Jerusalem, Israel}
\affil[2]{Department of Cognitive Sciences, The Federmann Center for the Study of Rationality, The Alexander Silberman Institute of Life Sciences, The Hebrew University, Jerusalem, Israel}

\affil[*]{tomer.barak@mail.huji.ac.il}

\begin{abstract}
When individuals encounter observations that violate their expectations, when will they adjust their expectations and when will they maintain them despite these observations? For example, when individuals expect objects of type A to be smaller than objects B, but observe the opposite, when will they adjust their expectation about the relationship between the two objects (to A being larger than B)? Naively, one would predict that the larger the violation, the greater the adaptation. However, experiments reveal that when violations are extreme, individuals are more likely to hold on to their prior expectations rather than adjust them. To address this puzzle, we tested the adaptation of artificial neural networks (ANNs) capable of relational learning and found a similar phenomenon: Standard learning dynamics dictates that small violations would lead to adjustments of expected relations while larger ones would be resolved using a different mechanism -- a change in object representation that bypasses the need for adaptation of the relational expectations. These results suggest that the experimentally-observed stability of prior expectations when facing large expectation violations is a natural consequence of learning dynamics and does not require any additional mechanisms. We conclude by discussing the effect of intermediate adaptation steps on this stability. 
\end{abstract}

\begin{document}

\flushbottom
\maketitle

\thispagestyle{empty}

\section*{Introduction}

Imagine strolling through an art museum, expecting awe-inspiring masterpieces. Suddenly, disrupting your expectations, you encounter, displayed in the vitrine, ... a banana. This can be framed as a violation of a relational expectation: On the one hand, the artistic value of museum displays is expected to be \textit{greater} than the artistic value of mundane objects. On the other hand, a banana seems to have a limited artistic value. There are two ways to resolve this inconsistency: recalibrate the expected relationship between the museum displays and mundane objects, or maintain this expectation and find an alternative explanation for the observation (e.g., find a deeper appreciation for the artistic value of bananas).  

Previous studies suggested the existence of distinct cognitive modules associated with the generation of representations and the encoding of relations both in humans and other species \cite{gentner_structure-mapping_1983, doumas_theory_2008, lazareva_multiple-pair_2008, lazareva_relational_2012, mansouri_emergence_2020, holyoak_relational_2021, miconi_neural_2025}. In the banana example, the \emph{representational} module, which extracts task-relevant features from inputs, determines the artistic value of objects, whether a Rodin sculpture or a banana. Adaptation of this module would correspond to finding merit in bananas. The \emph{relational} module encodes the expected relationship between representations, or between these representations and a predefined anchor. For instance, the relational module in the banana scenario encodes the expectation that objects displayed in an art museum surpass a certain threshold of artistic value, and its adaptation would result in decreasing this threshold. While in the banana example both modules can simultaneously adapt--slightly changing the representation of bananas and the expectation from museums--this is not the case in all violations of relational expectations. 

Consider a scientist who has consistently observed that particles of type B are larger than particles of type A. With new experimental techniques, experiments suggest the opposite: particle A is actually larger than particle B. This unexpected finding forces the scientist to choose between two alternative adaptation pathways: maintain the view that B is larger than A by, for example, questioning the validity of the experimental results (adapt the representational module), or alternatively, update their view and conclude that A is, indeed, larger than B (adapt the relational module). This scenario sets two distinct possible adaptation pathways.

Our study was motivated by recent experiments that found that the choice between resolution pathways exhibits an inverted U-shaped dependence on the size of the violation \cite{filipowicz_rejecting_2018, hird_boundary_2019, spicer_theory_2020, kube_how_2022}: When the violation is minute, its effect on the expectations is small. Larger violations have a larger effect on the expectations. Even larger violations (\textit{extreme} violations), however, fail to alter expectations. Somewhat similar findings were also observed in non-human animals in the framework of associative learning. When moderate, stronger unconditional stimuli elicit stronger conditional response. However, when extreme, the magnitude of the resultant conditional response \textit{decreases} with the magnitude of unconditional stimulus \cite{pavlov_lectures_1941, leaton_potentiated_1985, davis_conditioned_1978}. These results are surprising because naively, operant learning, predictive coding and Bayesian models assert that the larger the prediction error, the greater the adaptation \cite{rescorla_theory_1972, rao_predictive_1999, knill_bayesian_2004, tenenbaum_how_2011, gopnik_reconstructing_2012}. Therefore, gating mechanisms that modulate the magnitude of the adaptation \cite{spicer_theory_2020}, for example, ``immunization'' mechanisms that prevent adaptation in extreme violation settings have been proposed \cite{w_expectancies_2015, pinquart_why_2021, panitz_revised_2021, rief_using_2022}. 

In this paper we show that the inverted U-shaped dependence of adaptation on the size of the violation naturally emerges in artificial neural networks (ANNs) that are trained to identify relations using standard (gradient) learning. We first present the results from a behavioral perspective using deep networks. Then, we explain these results by mathematically analyzing a simplified model.

\section*{Results}

\subsection*{The order discrimination task}

\begin{figure}[H]
\begin{center}
\centerline{\includegraphics[width=0.85\linewidth]{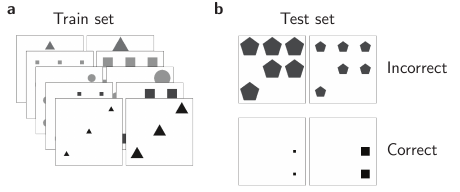}}
\caption{\textbf{Order discrimination task.} \textbf{(a)} Training set that demonstrates an underlying rule between images. The shapes in the right images are larger than those in the left images. The difference is characterized by $\alpha$. In these examples, $\alpha=0.5$, corresponding to half of the maximal size difference possible in our simulations. \textbf{(b)} The task is to determine whether two images are in the correct order according to the rule that characterizes the training set.}
\label{fig:task}
\end{center}
\end{figure}

We study adaptation using the order discrimination task: Agents are presented with image pairs and are instructed to determine, for each pair, if it was presented in the correct or reverse order. Each image depicts shapes arranged on a $3\times 3$ grid and is characterized by five features: the grayscale color of the shapes, their number, size, grid arrangement, and shape type (Fig. \ref{fig:task}). The first three features -- color, number, and size -- are described by a scalar number, establishing natural possible order relations between images. The ``correct'' order in the task depends on the identity of the relevant feature (color, size, or number), termed the \textit{predictive feature}, and whether this feature increases or decreases from left to right. 

To teach the underlying rule, the agents are presented with a series of ``correctly'' ordered image pairs (Fig. \ref{fig:task}a), such that the predictive feature changes according to the underlying rule, while the remaining features, irrelevant to deciding the correct order, are randomly chosen for each pair but remain constant within the pair (as in the test images). In the main text of this paper we present the results when the predictive feature was the size. We demonstrate the generality of our findings by presenting similar results in the Supplementary Information, when the predictive features were color or number (Figures S1-S4). 

A key parameter in this task is the difference in the predictive feature values between the two images, a quantity that we denote by $\alpha$. A positive $\alpha$ implies that the predictive feature increases from left to right, whereas a negative $\alpha$ implies that it decreases. The absolute value of $\alpha$ determines the magnitude of change: $\alpha=0$ implies that the feature does not change between the two images, whereas $\alpha=\pm1$ signifies maximal difference. In the real world, $\alpha$ would correspond to the (scaled) objective difference between objects. In the example of the scientist measuring particle sizes, $\alpha$ is the (scaled) objective difference between the sizes of particle $A$ and particle $B$.

\subsection*{The ANN}

As agents, we used ANNs that were designed to emulate relational learning \cite{santoro_simple_2017, barrett_measuring_2018, sung_learning_2018, hill_learning_2018}. Specifically, our networks were comprised of two modules. The first is a representational module, an encoder which we denote by $Z$. Its goal is to extract a relevant feature from inputs. For each of the two images of a pair, $\*{x}$ and $\*{x'}$ (left image and right image, respectively), it maps the $n\times n$ image to one-dimensional variables $Z_{\*w}(\*{x})$ and $Z_{\*w}(\*{x'})$, where $\*w$ are trainable parameters. We implemented this mapping using a multilayered convolutional network. A similar architecture was shown in a previous study to be able to extract the relevant features in an intelligence test \cite{barak_untrained_2024}. The difference between the representations of the two images is given by $\Delta Z=Z_{\*w}(\*{x'})-Z_{\*w}(\*{x})$. 

The relational module, which we denote as $R_\theta$, characterizes the expected relation between the representations of the two images of the pair. In general, a relational module could be any function $R_\theta\left(Z_{\*{w}}\left(\*{x}'\right),Z_{\*{w}}\left(\*{x}\right)\right)$. We used a constant, single-parameter function that encodes the expected difference between representations, $R_\theta=\theta$.

Formally, we define a loss function for a pair of images as:
\begin{equation}
\label{eq:loss}
\mathcal{L}(\*{w},\theta) =\left(\left(Z_{\*{w}}\left(\*{x}'\right) - Z_{\*{w}}\left(\*{x}\right)\right) - R_\theta\right)^2 =  \left(\Delta Z - \theta\right)^2 .
\end{equation}
If $\Delta Z=\theta$, that is, if the difference between the two representations $\Delta Z$ is equal to the expected relationship $\theta$ then the loss function is minimized. Formulated this way, however, if $Z_{\*w}(\*{x})=0$ for all $\*{x}$ then $\Delta Z=0$ for all $\*{x}$ and $\theta=0$ will trivially minimize the loss. To avoid the model collapsing into this trivial solution, we defined a regularized loss function, in which $\Delta Z$ and $\theta$ are driven to reside on a ring,
\begin{equation}
\label{eq:loss_reg}
\tilde{\mathcal{L}}(\*{w},\theta) = \mathcal{L}  + \lambda\left(\Delta Z^2 + \theta^2 - r^2 \right)^2.
\end{equation}
$\lambda>0$ and $r$ are hyper-parameters. With this additional regularization term, solutions such that $\tilde{\mathcal{L}}(\*{w},\theta)=0$, if attainable, would reside in the points where the line $\Delta Z=\theta$ intersects with the ring $\Delta Z^2=\theta^2=r^2/2$, $\Delta Z=\theta= r/\sqrt{2}$ and $\Delta Z=\theta=- r/\sqrt{2}$.

\begin{figure}[H]
\begin{center}
\centerline{\includegraphics[width=0.9\linewidth]{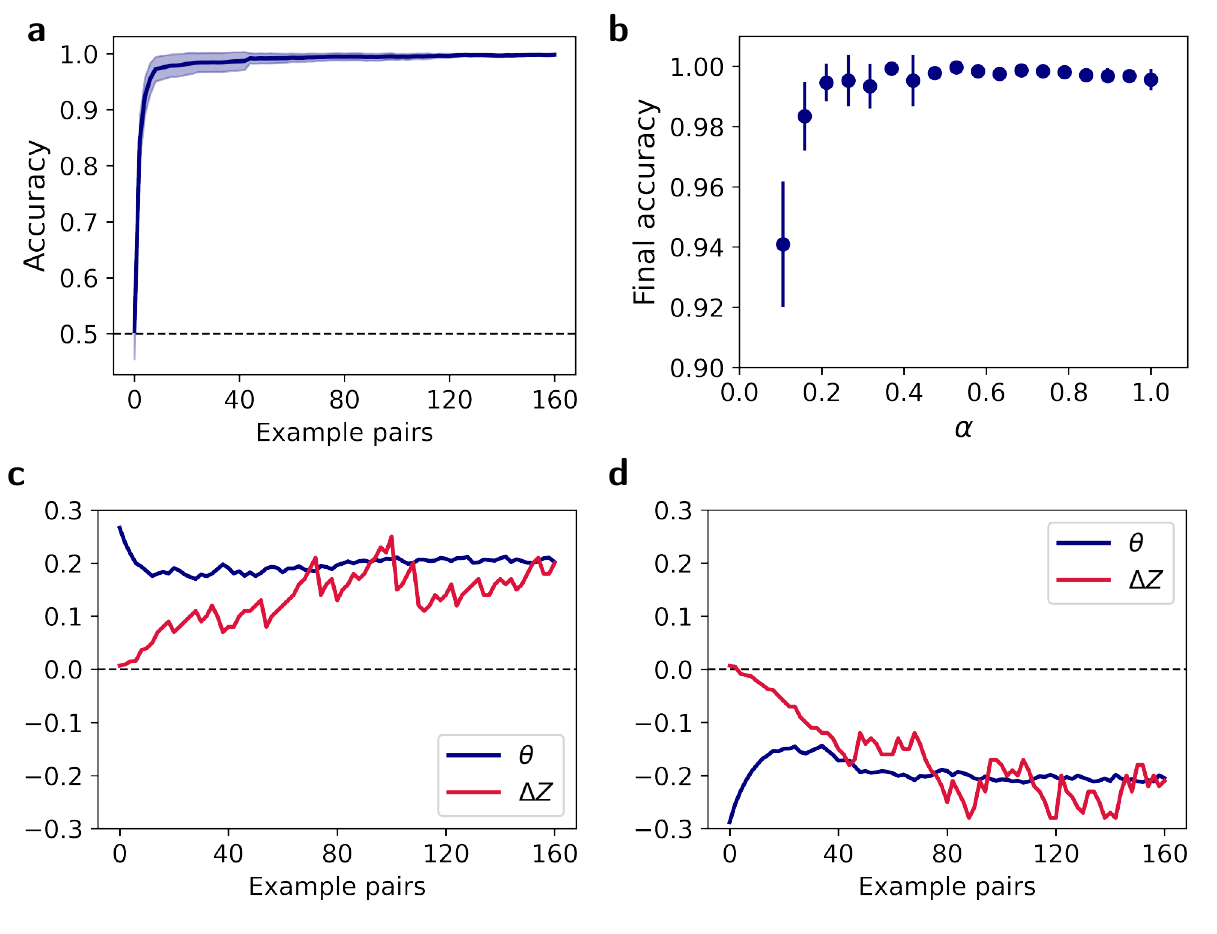}}
\caption{\textbf{Task performance and solutions.} (\textbf{a}) The average test accuracy of 100 networks trained on a task where the predictive feature, size, changed by $\alpha=0.5$. (\textbf{b}) The final test accuracies for various values of $\alpha$, averaged over 100 networks per $\alpha$. Error shades and bars correspond to $95\%$ CI. (\textbf{c-d}) The model can solve the task using two different internal strategies. In one strategy, \textbf{c}, a representative network learned to measure "largeness," where it perceived the shapes getting larger (positive $\Delta Z$) and expected them to get larger (positive $\theta$). In an equally effective strategy, \textbf{d}, another network learned to measure "smallness," where it perceived the shapes becoming less small (negative $\Delta Z$) and expected "smallness" to decrease (negative $\theta$). Both solutions are equally valid for solving the task.}
\label{fig:task_eval}
\end{center}
\end{figure}

The model parameters $\*{w}$ and $\theta$ are trainable and we used Stochastic Gradient Descent (SGD) on $\tilde{\mathcal{L}}(\*{w},\theta)$ to learn them (see Methods). To evaluate the performance of the ANN, we measured its ability to determine the order of novel test images. Specifically, we presented the ANN with two images, measured the values of the loss function associated with the two possible orders of these images (left-right or right-left), and chose the order that minimized the loss function. Fig. \ref{fig:task_eval}a, depicts the average performance of 100 ANNs as a function of the size of the training set when $\alpha=0.5$, showing that the networks successfully learned to solve the task after less than $40$ image pairs. The high performance holds for other values of $\alpha$. We tested $\alpha$ values ranging from $0.1$ to $1$, training them on $160$ image pairs, and found that throughout this range, the ANNs performance exceeded $90\%$ accuracy (Fig. \ref{fig:task_eval}b). Larger differences in the predicted feature between the two images ($\alpha$) were associated with higher performance, signifying that the magnitude of $\alpha$ is a measure of the difficulty of the task.  This robust learning performance, in a model with distinct representational and relational modules, reflects the ability shown in humans and non-human animals to learn relationships regardless of absolute attributes. Moreover, the higher performance with larger $\alpha$ values align with observations from those studies, where clear distinctions between stimuli support more robust relational behavior \cite{lazareva_multiple-pair_2008, lazareva_relational_2012, ribes-inesta_comparison_2020, leon_rbdt_2021}.

It is instructive to separately consider how each of the two modules adapt in the process of learning. This is depicted in Fig. \ref{fig:task_eval}c for one representative network, where we plot the values of $\Delta Z$ and $\theta$ as a function of example pairs. According to Eq. \eqref{eq:loss_reg}, the regularized loss is minimized when $\Delta Z^2 = \theta^2 = r^2/2$. In this simulation, $r^2=0.1$. While the values of $\Delta Z$ and $\theta$ vary between example pairs, they approach $\Delta Z\approx\theta\approx r/\sqrt 2 \approx 0.2$. Fig. \ref{fig:task_eval}d depicts another example network. In this simulation, which differed only in the initial parameters, $\Delta Z$ and $\theta$ converged to a negative solution, where $\Delta Z\approx\theta\approx -r/\sqrt 2 \approx -0.2$. From a point of view of task performance, the positive and negative solutions ($\pm r/\sqrt{2}$) are identical. In the positive solution, the representational module learns to extract the size of the shapes, that is, how large they are, and the relational module learns that this size increases between the two images (from left to right). In the negative solution, the representational module learns to extract the ``smallness'' of the shapes (the negative of the size) and the relational module learns that the ``smallness'' decreases between the two images.

\subsection*{Dual adaptation pathways in ANNs}

To study the violation of relational expectations in the ANNs, we trained them with sequences of image pairs, characterized by a specific predictive feature that changes by $\alpha_1>0$ between the two images of the pair. After learning, we changed the training set's relation rule to $-\alpha_2$ where $\alpha_2>0$. That is, the shapes' sizes decreased rather than increased from left to right, violating the relational expectation. We continued training the ANNs with the new rule and measured the performance of the ANNs as a function of examples.

The rule sizes before and after reversal, $\alpha_1$ and $\alpha_2$, determine the magnitude of the violation. Specifically, we expect that large $\alpha_1$ and $\alpha_2$ would correspond to a large violation while small $\alpha_1$ and $\alpha_2$ correspond to a small violation. To illustrate the core phenomenon, we begin our analysis by considering the specific case of a symmetric rule reversal: $\alpha_1=\alpha_2$. To simplify notations, we write $\alpha_1=\alpha$ and $\alpha_2=\alpha$, therefore analyzing the case of rule reversal, $\alpha\to-\alpha$. Later, we study how $\alpha_1$ and $\alpha_2$ independently affect the adaptation pathway.

\begin{figure}[H]
\begin{center}
\centerline{\includegraphics[width=0.9\linewidth]{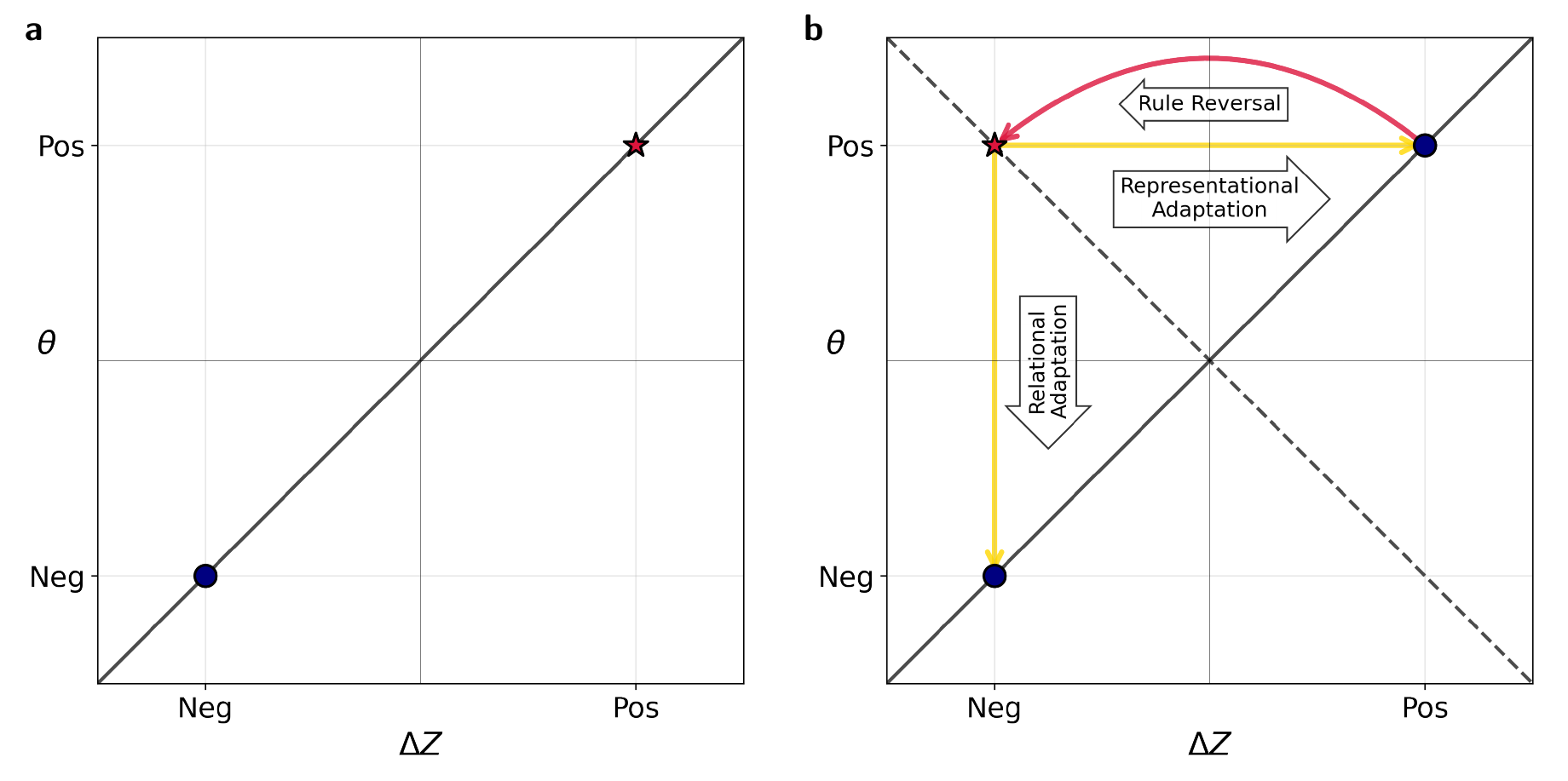}}
\caption{\textbf{Illustration of the two adaptation pathways.} (\textbf{a}) The two equivalent solutions for the relational task. One solution is to encode the size of the objects and expect that the size increases (positive $\Delta Z$ and $\theta$, marked by a star). In the other solution, the smallness decreases (negative $\Delta Z$ and $\theta$). The diagonal line all solutions in which $\Delta Z=\theta$. (\textbf{b}) Adaptation pathways following rule reversal. Without loss of generality, we consider an agent that has learned the positive solution. After learning the initial rule, the rule is reversed, flipping the sign of $\Delta Z$, making it inconsistent with positive $\theta$ (marked by a star). There are two adaptation pathways: \textit{relational adaptation} - maintaining the sign of $\Delta Z$ but reversing the sign of the expectation $\theta$; or \textit{representational adaptation} - changing the sign of $\Delta Z$ to encode the smallness of objects, while maintaining the expectation that the relevant feature increases ($\theta>0$). The dashed diagonal line represent expectations that are opposite to the observations ($\theta = -\Delta Z$).}
\label{fig:model_schema}
\end{center}
\end{figure}

Crucially, there are two ways of resolving a rule reversal violation. Recall that in Fig. \ref{fig:task_eval} we saw two different solutions that networks trained on the task identified. In one, both $\Delta Z$ and $\theta$ were positive, while in the other, both were negative. There, we discussed the fact that the solution that minimizes the loss function is not unique: $\Delta Z = \theta = r/\sqrt 2$ and $\Delta Z = \theta = -r/\sqrt 2$ both minimize the loss function. These two solutions are depicted schematically in Fig. \ref{fig:model_schema}a. Following rule reversal, $\Delta Z$ necessarily changes its sign, violating the expectation set by $\theta$. As illustrated in Fig. \ref{fig:model_schema}b, one possibility for resolving the expectation violation is by changing the weights of the representational module, $Z_\*{w}$, so that $\Delta Z$ would return to its pre-reversal sign keeping the relational module unchanged. The other possibility is that the representational module retains its post-reversal sign of $\Delta Z$ while the relational module $\theta$ changes its sign. Both can lead to exactly the same level of performance. We hypothesized that magnitude of $\alpha$ would determine which of the two adaptation pathways would be taken by the ANNs.

\begin{figure}[H]
\begin{center}
\centerline{\includegraphics[width=\linewidth]{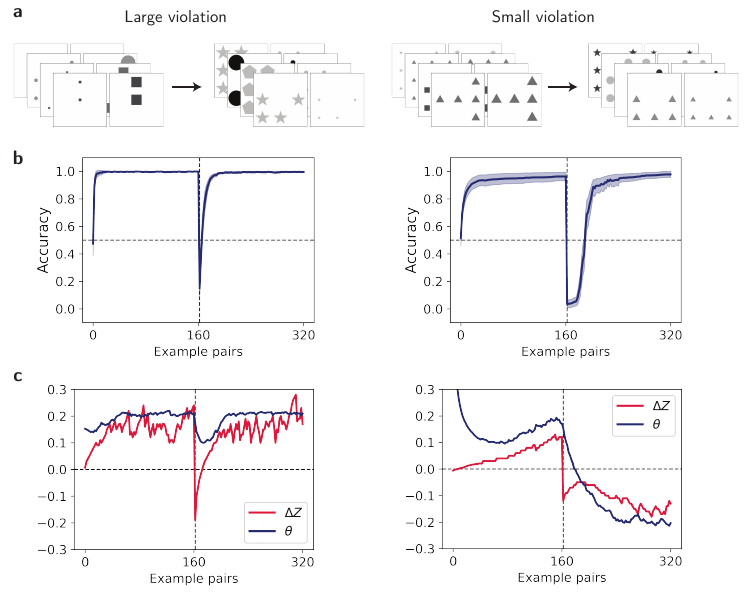}}
\caption{\textbf{Dual adaptation pathways following rule reversal.} Demonstration of the two adaptation pathways, depending on the violation magnitude, when the rule changes from $\alpha$ to $-\alpha$. (a) Example image pairs before and after the rule reversal for a large violation ($\alpha = 0.8$, left) and a small violation ($\alpha = 0.2$, right). (b) Performance across 100 networks as a function of examples (shaded regions: 95\% CI). (c) Representative demonstration of the evolution of $\Delta Z$ (red) and $\theta$ (blue) for each violation magnitude. Initially, both networks converge to the positive solution ($\theta, \Delta Z > 0$). After reversal, $\Delta Z$ flips immediately due to the change in the image order. Optimization drives $\Delta Z$ and $\theta$ towards each other, approaching $\Delta Z=0$ and $\theta =0$. For large violations (left), the representation module "wins" and adaptation restores $\Delta Z$ while keeping $\theta$ unchanged. For small violations (right), adaptation eventually occurs via the relational module, flipping the sign of $\theta$ to match the negative $\Delta Z$.}
\label{fig:task_switch}
\end{center}
\end{figure}

Going back to the ANNs, we first considered two violation magnitudes: a larger violation ($\alpha=0.8$, Fig. \ref{fig:task_switch}a left) and a smaller violation ($\alpha=0.2$, Fig. \ref{fig:task_switch}a right). We measured the ANNs performance as a function of examples before and after reversing the rule (Fig. \ref{fig:task_switch}b). In both the larger and the smaller violation conditions, performance levels just before the sign reversal (image pair $160$) were almost perfect. The first image pairs immediately following the reversal were almost always incorrectly ordered by the ANNs (performance level close to $0$). With examples, however, the networks adapted to the new rule, achieving almost perfect performance after additional $160$ image pairs. Notably, learning was slower for the more difficult task associated with the smaller value of $\alpha$. Also, adaptation to the reversal was slower than initial learning, a phenomenon that has been previously reported in human learning \cite{monsell_task_2003}.

To dissect the roles of the two modules in this reversal adaptation, Fig. \ref{fig:task_switch}c depicts the values of $\Delta Z$ (red) and $\theta$ (blue) in representative networks adapting to the large and small violations. Before reversal, both networks converged to comparable values of $\Delta Z$ and $\theta$. Immediately after the rule reversal, $\Delta Z$ flipped. This is because the image order was reversed -- the sizes of the shapes decreased, rather than increased between the two images -- and the representation module reflected it. The value of $\theta$, however, remained unchanged. This is because the network ``expected'' the sizes of the shapes to increase rather than decrease. With training, the system resolves this violation.

In our simulations, we found that when the violation was large (Fig. \ref{fig:task_switch}c left), the representational module $Z$ adapted so that $\Delta Z$ returned to its pre-reversal values. By contrast, when the violation was small (Fig. \ref{fig:task_switch}c right), $\Delta Z$ remained negative and the violation was resolved by a change in the sign of the relational module $\theta$.

\begin{figure}[H]
 \begin{center}
 \centerline{\includegraphics[width=0.9\linewidth]{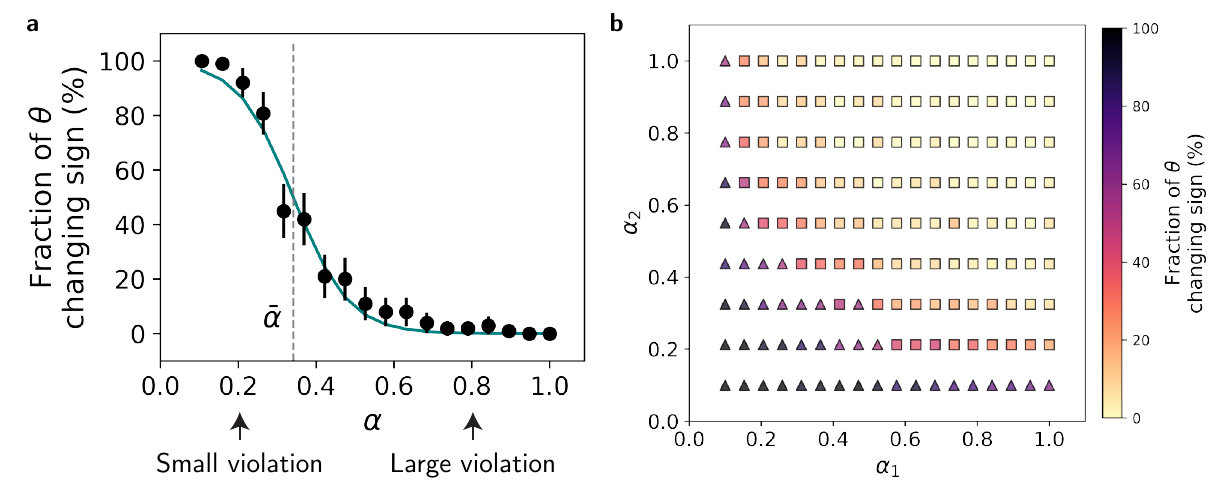}}
\caption{\textbf{Adaptation pathway dependence on violation magnitude.} (\textbf{a}) Fraction of networks ($n=100$) adapting via relational module ($\theta$ sign change) for rule reversal $\alpha\to-\alpha$, fitted with logistic function (green) to estimate inflection point $\bar{\alpha}$.Arrows mark the $\alpha$ values from Fig. \ref{fig:task_switch}. The logistic function fit has two parameters, the inflection point, $\bar\alpha\sim 0.34$, and the slope parameter, $\sim 14.10$ (see Methods). Error bars: 95\% CI. (\textbf{b}) Adaptation pathways for general rule changes $\alpha_1\to-\alpha_2$. Color indicates fraction (out of $n=50$) adapting via relational module ($\theta$). Squares: representational module ($Z$) dominant (ratio $<50\%$); triangles: relational module dominant (ratio $\geq 50\%$).}
\label{fig:concept_ada_ratio}
 \end{center}
 \end{figure}

Are these results representative? We simulated reversal adaptation for different values of $\alpha$, each time simulating $100$ randomly-initialized networks, and measuring the fraction of times in which the adaptation involved a change in the sign of $\theta$, which indicates that the relational module dominates the adaptation. In line with the examples of Fig. \ref{fig:task_switch}c, the smaller $\alpha$, the larger was the fraction of networks in which the relational module flipped its sign in response to the rule reversal (Fig. \ref{fig:concept_ada_ratio}a). The transition between the two adaptation pathways, an inflection point marked by $\bar\alpha$, was at $0.34\pm0.02$. These results demonstrate that large violations ($\alpha>\bar\alpha$) can inhibit adjustments to relational expectations, leading to adaptation in object representations instead.

The results of this section are reminiscent of the surprising part of the experimentally-observed inverted U-shaped dependence of relational adaptation to the size of the violation discussed in the Introduction. When the violation is modest ($\alpha$ is small), the relational expectation adapts: $\theta$ changes its sign when the order of images is reversed. By contrast, it remains unchanged when the violation is large. Instead, the violation is resolved by the network changing its representation. This behavior does not require any explicit immunization mechanism. Rather, it naturally emerges from the dynamics of learning.

\subsection*{General rule reversals}

In the analysis above, we considered the special case of symmetric rule reversal, in which $\alpha_1=\alpha_2$. Now, we study the specific contribution of these two parameters to the choice of an adaptation pathway. To that end, we studied the adaptation of ANNs to a different pairs of $\alpha_1$ and $\alpha_2$, as depicted in Fig. \ref{fig:concept_ada_ratio}b. For each pair $\left(\alpha_1,\alpha_2\right)$ we measured the fraction of networks in which adaptation was associated with a change in the sign of $\theta$ (color coded). To better visualize the transition point (which was denoted by $\bar \alpha$ in the case of $\alpha_1=\alpha_2$), the symbol (square vs. triangle) denotes whether this fraction was smaller or larger than $50\%$. These simulations show that the adaptation pathway depends both on $\alpha_1$ and $\alpha_2$. The larger $\alpha_1$ and the larger $\alpha_2$, the more likely it is that the representational module will change its sign. However, a large $\alpha_1$ can compensate for by a small $\alpha_2$ and vice versa. The boundary between the two adaptation pathways resembles a hyperbolic curve. This is not a coincidence. Below we prove, in a simplified model that in the limit of weak regularization, the boundary is, indeed, a hyperbolic function in the $\alpha_1\times \alpha_2$ plane.

Next we study the implications regarding our ability to use shaping to facilitate or inhibit adaptation of relational expectations. 

\subsection*{Shaping adaptation through an intermediate rule}

Understanding how individuals adapt to violations of their expectations is important for clinical psychology, as expectation persistence and change are central to mental health interventions. Clinical research has shown that maladaptive expectations contribute to disorders such as anxiety and depression, where individuals often maintain dysfunctional expectations even in the face of disconfirming evidence \cite{rief_using_2022}. Effective psychological treatments leverage expectation violations to induce cognitive and behavioral change. Yet, while moderate expectation violations are most effective in altering beliefs, extreme violations risk reinforcing rigid mental models, preventing adaptation \cite{spicer_theory_2020, rief_using_2022}. In this section we show that adding an intermediate rule in the reversal task can alter the adaptation pathway, thereby steering the adaptation process toward a preferred adaptation strategy. 

We compared adaptation in reversal task in which the rule is reversed in one step, as before ($\alpha \to -\alpha$) to adaptation when our agents also adapt to an intermediate step ($\alpha \to \beta \to -\alpha$). We hypothesized that the value of $\bar\alpha$ would inversely depend on the magnitude of the intermediate step: larger intermediate rules would lower $\bar\alpha$, making relational adaptation less likely, whereas smaller intermediate steps would increase $\bar\alpha$, favoring relational adaptation. The intuition behind this hypothesis is that the relational module can change its sign only when the rule reverses. Therefore, when $\beta>0$, if $\beta>\alpha$ then the intermediate step enhances the violation (increases $\alpha_1$ of Fig. \ref{fig:concept_ada_ratio}b) and therefore decreases the probability of a relational adaptation. By contrast, $\beta<\alpha$ decreases the magnitude of the violation and therefore, increases the probability of a relational adaptation. The effect of a negative $\beta$ is similar. This time, the focus is on the transition ($\alpha \to \beta$), where $-\beta$ takes the role of $\alpha_2$ of Figure \ref{fig:concept_ada_ratio}b. 

\begin{figure}[H]
\begin{center}
\centerline{\includegraphics[width=\linewidth]{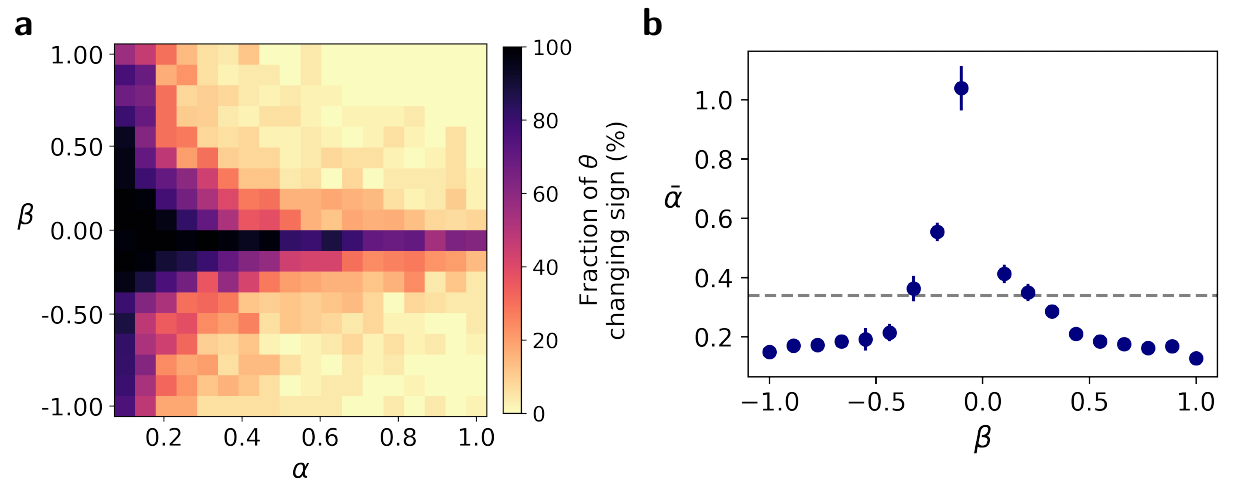}}
\caption{\textbf{Intermediate learning step: influence on adaptation pathways.} The threshold between relational and representational adaptation ($\bar\alpha$) is modulated by the magnitude of an intermediate step $\beta$. (\textbf{a}) The fraction of networks that adapt $\theta$ for training with an intermediate step $\alpha\to\beta\to-\alpha$ for various pairs of $(\alpha,\beta)$. (\textbf{b}) Large values of $|\beta|$ lower $\bar\alpha$, promoting representational adaptation, while small $|\beta|$ increases $\bar\alpha$, favoring relational adaptation. The gray dashed line represents $\bar\alpha$ in the absence of an intermediate step, highlighting how structured transitions can shape the adaptation process.}
\label{fig:ANNs_betas}
\end{center}
\end{figure}

To test this hypothesis, we trained ANNs using the $\alpha \to \beta \to -\alpha$ paradigm, using $160$ image pairs for each rule (total $160\times3=480$ image pairs). For each pair $\left(\alpha,\beta\right)$ we trained $50$ ANNs and computed the fraction of networks in which the sign of $\theta$ changed from image pair $160$ (after the network learned the $\alpha$ rule) to image pair $480$ (after the network learned the $-\alpha$ rule). The results are depicted in Fig. \ref{fig:ANNs_betas}a. Then, for each value of $\alpha$ and $\beta$, we computed $\bar \alpha$ by fitting a logistic function to the computed fraction as a function of $\alpha$, as in Fig. \ref{fig:concept_ada_ratio}a. The values of $\bar \alpha$ as a function of $\beta$ are presented in Fig. \ref{fig:ANNs_betas}b. Indeed, for large values of $|\beta|$, $\bar\alpha$ was smaller than that computed in the  $\alpha \to -\alpha$ paradigm (dashed horizontal line, computed in Fig. \ref{fig:concept_ada_ratio}a), while small values of $|\beta|$ increased $\bar\alpha$. These findings demonstrate that adaptation pathways can be influenced by the sequence of changes between them. By strategically introducing an intermediate step, we can shift the boundary between relational and representational adaptation, effectively shaping the learning process. Specifically, small values of $\beta$ increase the probability that the relational network would adapt. 

\subsection*{Simplified model analysis}

\subsubsection*{Setting up the model and loss function}

To gain an analytical understanding as to why small and large violations lead to qualitatively different adaptation mechanisms, and the extent to which these results depend on the particularities of the model that we studied, we considered adaptation to violation in a simplified model, in which the pair of images was replaced by a pair of scalars, denoting the predictive features in the pair of images. That is, $\*x = x$ and $\*x' = x'$ and their relation is their difference, $x'-x=\alpha$. 

Now, that the feature is explicitly provided to the agent, we model the representational module using a single-weight linear encoder, $Z_\*{w}(x)=wx$. Under this formulation, the difference in the representations of the pair of stimuli is simply $\Delta Z =wx'-wx =w \alpha$. The corresponding loss function then becomes
\begin{equation}
\label{eq:loss2}
    \tilde{\mathcal{L}}(w,\theta) = \left( w \alpha -\theta \right)^2 + \lambda \left( \left(w \alpha \right)^2 + \theta^2 -r^2 \right)^2.
\end{equation}

This loss function captures two key components: (1) the squared error between the representational difference $w\alpha$ and the relational expectation $\theta$, and (2) a regularization term ensuring that solutions remain on a ring.

\subsubsection*{SGD dynamics and differential equations}

The advantage of this simplified formulation is that we can now use mathematical techniques, borrowed from the field of non-linear dynamics, to analytically characterize the adaptation dynamics \cite{strogatz_nonlinear_2018}. In the limit of an infinitesimally small learning rate, the SGD dynamics that acts to minimize the loss function can be expressed as a set of differential equations governing the evolution in time (a proxy of example pairs) of $w$ and $\theta$ \cite{benaim_dynamics_1999}:
\begin{equation}
\begin{split}
   \dot{w} &= - \alpha\left( w \alpha  - \theta \right) - 2\lambda \alpha^2 \left( \left(w \alpha \right)^2  + \theta^2 -r^2 \right) w   \\
    \dot{\theta} &=  \left( w \alpha - \theta \right) - 2\lambda \left( \left(w \alpha \right)^2 + \theta^2 -r^2 \right)\theta.
\end{split}
\end{equation}

The evolution of $w$ and $\theta$ over time represents the adaptation of these parameters through learning with successive example pairs. Thus, by analyzing the temporal dynamics of $w$ and $\theta$, we can gain insight into the adaptation pathway of the system. Instead of directly examining the dynamics of $w$, it is more instructive to focus on the dynamics of the output of the representational module, $\Delta Z = w \alpha$. Substituting this into the differential equations, we obtain the following system of equations:
\begin{equation}
\label{eq:dynamics}
\begin{split}
    \frac{1}{\alpha^2}\dot{\Delta Z} &=  \left(  \theta - \Delta Z \right) - 2\lambda \left( \Delta Z ^2 + \theta^2 -r^2 \right)\Delta Z    \\
    \dot{\theta} &=  \left( \Delta Z - \theta \right) - 2\lambda \left( \Delta Z  ^2 + \theta^2 -r^2 \right)\theta.
\end{split}
\end{equation}


Recall that when studying the adaptation of the ANN to the image pairs, we identified two solutions that minimize the loss: $\Delta Z=\theta=\pm r/\sqrt 2$. The two solutions are associated with the same level of performance, they differ in how the system encodes the relationship: when $\Delta Z=\theta=r/\sqrt 2$, the predictive feature increases, whereas when $\Delta Z = \theta = -r/\sqrt 2$, it decreases, as discussed in Fig. \ref{fig:task_eval}.

In gradient-based systems, learning dynamics converge to a stable fixed point where $\dot{\Delta Z}=\dot{\theta}=0$. In the Methods section we prove that $\Delta Z=\theta=\pm r/\sqrt 2$ are the only stable fixed points of the dynamics, Eq. \eqref{eq:dynamics}. Thus in general, the dynamics would converge to either the positive or the negative fixed point.

\subsubsection*{Adaptation pathways following rule reversal}

To investigate the dynamics following rule reversal, we consider a system that has converged to the positive fixed point $\Delta Z=\theta=r/\sqrt 2$. Following the reversal of the rule, the sign of the representational difference $\Delta Z$ flips to $-r/\sqrt{2}$. As a result, the value of $\left(\Delta Z - \theta \right)^2$ in the loss becomes non-zero, reflecting the violation of the expected relationship.

At this point, the gradient dynamics would resolve the violation by either driving the system back to the positive fixed point $\Delta Z=\theta=r/\sqrt 2$, adapting the representational module, or to the negative fixed point $\Delta Z=\theta=-r/\sqrt 2$, which adapts the relational expectation (as illustrated in Figure \ref{fig:model_schema}).

To understand how the magnitude of $\alpha$ affects this adaptation pathway, we first considered the dynamics of a weakly regularized system, where $\lambda\ll1$. In this case, the dynamics first minimize the unregularized part of the loss, $\left(\Delta Z - \theta \right)^2$, driving the system to $\Delta Z = \theta$, and then the regularization kicks in to set the system on the ring $\Delta Z^2=\theta^2=r^2/2$. We show below that this sequential adaptation pattern -- first aligning $\Delta Z$ and $\theta$, then enforcing the regularization constraint -- provides a key insight into the behavior of the more complex ANN.

Without regularization, the dynamical equations simplify to
\begin{equation}
\label{eq:dynamics_linear}
\begin{split}
    \dot{\Delta Z} &=  -\alpha^2 \left( \Delta Z -  \theta \right)  \\
    \dot{\theta} &=  \left( \Delta Z - \theta \right).
\end{split}
\end{equation} 

When $\alpha>1$, the dynamics of $\Delta Z$ is faster than that of $\theta$ (because of the $\alpha^2$ prefactor). Consequently, adaptation is expected to be dominated by a change in the sign of $\Delta Z$. By contrast, when $\alpha<1$, adaptation is expected to rely on a change in $\theta$. This can be shown more formally. In the Methods section we show that the dynamics converge to 
\begin{equation}\label{eq:dynamics_linear_general_fp}
   \Delta Z = \theta =  \frac{\alpha^2 \theta (0)+ \Delta Z (0)}{\alpha^2+1},
\end{equation}
where $\Delta Z (0)$ and $\theta (0)$ denote the values of $\Delta Z$ and $\theta$ before the reversal. 

Substituting the initial conditions $\Delta Z=-r/\sqrt 2$ and $\theta=r/\sqrt 2$, we find that after reversal, $\theta$ will change its sign if and only if $\alpha<1$. 

Considering now the full model (Eq. \eqref{eq:dynamics}), we identify the following symmetry: Immediately after reversal $-\Delta Z\left(0\right)=\theta\left(0\right)$. Considering the equations for $-\dot{\Delta Z}$ and $\dot \theta$, they are also symmetric to swapping $-\Delta Z$ and $\theta$, as long as $\alpha^2$ is replaced by $1/\alpha^2$. Consequently, if for a particular value of $\alpha'$, $\Delta Z$ would change its sign in the reversal protocol, it would be $\theta$ which changes its sign if $1/\alpha'$ is used, indicating that also in the full model, $\alpha=1$ is the transition point between the two, qualitatively-different modes of adaptation.  

\begin{figure}[H]
\begin{center}
\centerline{\includegraphics[width=0.9\linewidth]{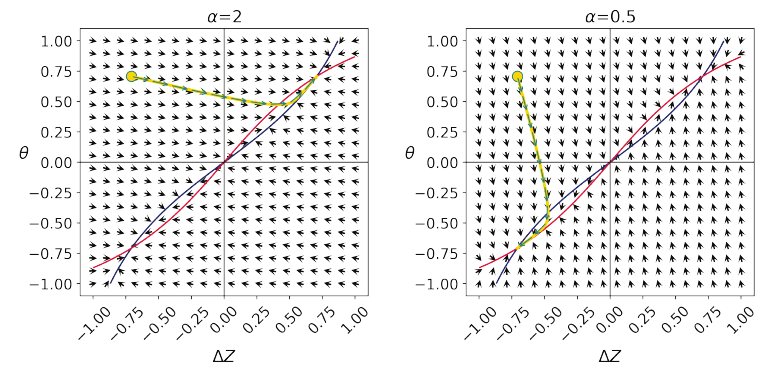}}
\caption{\textbf{The adaptation dynamics in the $\Delta Z \times \theta$ plane.}
Visualization of the two adaptation pathways as two trajectories in the simplified model's internal state represented by $\Delta Z$ and $\theta$. The arrows represent the "downhill" direction for learning: the opposite direction of the loss gradient with respect to $\Delta Z$ and $\theta$. The blue and red nullclines represent the loci where $\dot{\Delta Z}=0$ and $\dot{\theta}=0$, respectively. Their intersections correspond to fixed points in which learning halts. They intersect at the stable fixed points at $\Delta Z=\theta=\pm r/\sqrt{2}$, and at the unstable fixed point $\Delta Z=\theta=0$. Before reversing the rule, the network start with positive $\Delta Z$ and $\theta$. Following the reversal of the rule, $\Delta Z$ flips it's sign (yellow circles). Adaptation when $\alpha$ is large (left panel) leads to reversing the sign $\Delta Z$, returning to the positive fixed point, whereas when $\alpha$ is small (right panel) it is $\theta$ that changes its sign to match the negative $\Delta Z$.}
\label{fig:quiver}
\end{center}
\end{figure}

Fig. \ref{fig:quiver} depicts these dynamics in the reversal paradigm using a phase portrait for two values of $\alpha$. When $\alpha$ is large (left) adaptation is dominated by a change in $\Delta Z$. The opposite, a dynamics that is dominated by a change of $\theta$ manifests when $\alpha$ is small (right). Notably, in simulating the model, we used a relatively weak regularization, $\lambda=0.1$. As a result, initially, the dynamics drive the system to the line $\Delta Z = \theta$, minimizing the unregularized term by either changing the sign of $\Delta Z$ or $\theta$, depending on the size of the violation. Then, when $\Delta Z \approx \theta$, the regularization term pushes the system towards one of the two fixed points, where $\Delta Z^{2}=\theta^2=r^2/2$.

\subsubsection*{The more general case of $\alpha_1 \to -\alpha_2$}

\begin{figure}[H]
 \begin{center}
\centerline{\includegraphics[width=0.85\linewidth]{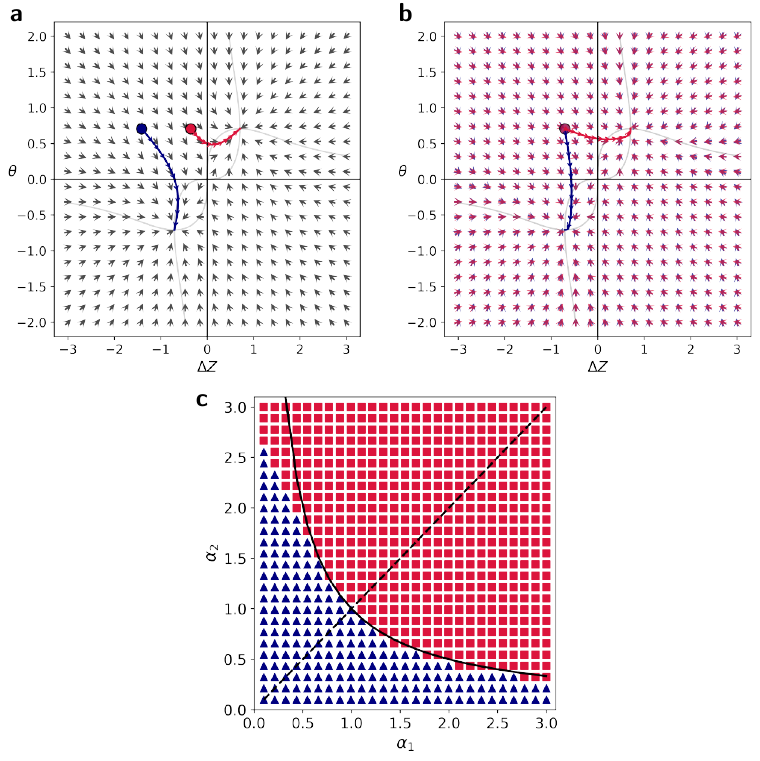}}
\caption{\textbf{How initial and new rule strengths determine the adaptation pathway.} This figure explores how the network adapts when the rule changes from an initial strength ($\alpha_{1}$) to an opposite rule of a different strength ($-\alpha_{2}$). The choice of adaptation pathway can be understood as a "race" between the representational module ($\Delta Z$) and the relational module ($\theta$). Two factors determine the winner. First, as illustrated in panel \textbf{a}, the ratio of the new rule's strength to the old one ($\alpha_{2}/\alpha_{1}$) determines the "starting position" for the race (here, $\alpha_2$ is fixed at $1$, while $\alpha_1$ is $2$ for the blue trajectory and $0.5$ for the red). Second, as shown in panel \textbf{b}, the strength of the new rule ($\alpha_{2}$) influences the relative "speed" of adaptation, changing the direction of the learning process (we used $\alpha_2=0.5$ for the blue trajectory and $\alpha_2=2$ for the red, while fixing the starting points by setting $\alpha_1=\alpha_2$). (\textbf{c}) The final race outcome for various combinations of $\alpha_{1}$ and $\alpha_{2}$. The boundary between the two pathways is determined by the combined strength of the rules. Specifically, the line $\alpha_{1}\alpha_{2}=1$ (black line) is a good fit for this boundary.}
\label{fig:Lin_a1_to_a2_explain}
 \end{center}
\vskip -0.2in
 \end{figure}

We also studied the system's dynamics in the more general adaptation of $\alpha_1\to -\alpha_2$. In this case, the initial state $\Delta Z(0)$ would generally not be $-r/\sqrt 2$ like in the symmetric rule reversal. Instead, it would be $\Delta Z (0) = -\frac{r}{\sqrt 2}\frac{\alpha_2}{\alpha_1}$ (see Methods). This is while $\theta(0)$ remains the same. Therefore, the ratio $\alpha_2/\alpha_1$ adjusts the distance of the system from the ordinates (y-axis), whose crossing corresponds to changing the sign of $\Delta Z$. The larger $\alpha_2/\alpha_1$, the longer is the path required for a change in the sign of $\Delta Z$ (Fig. \ref{fig:Lin_a1_to_a2_explain}a). Another modification to the dynamics is that $\alpha$ in Eq. \eqref{eq:dynamics_linear}, which sets the gradient of $\Delta Z$ compared with those of $\theta$, is replaced by $\alpha_2$. Therefore, the value of $\alpha_2$ determines the direction of the gradient, or the relative speed of adaptation of the two modules: The larger $\alpha_2$, the more the gradient is aligned with $\Delta Z$ compared to $\theta$ (more horizontal) (Fig. \ref{fig:Lin_a1_to_a2_explain}b).

Together, we can view the dynamics as a ``race'' between the two adaptation pathways. The ratio $\frac{\alpha_2}{\alpha_1}$ determines the starting points of the two adaptation pathways, and hence the relative distance to the finish line. The larger this ratio, the smaller is the relative distance required for adaptation by the relation module. The larger $\alpha_2$, the faster is the dynamics of $\Delta Z$, compared to that of $\theta$. Together, the relation module is more likely to adapt when the ratio $\frac{\alpha_2}{\alpha_1}$ is large and when  $\alpha_2$ is small (Fig. \ref{fig:Lin_a1_to_a2_explain}c). 

As discussed above, when regularization is weak, we can analytically predict the ``winner'' from the ratio between the relative distances of $\Delta Z$ and $\theta$ to the finish line, $\alpha_2/\alpha_1$, and the ratio between their speeds, $\alpha_2^2$. If $\alpha_1\alpha_2>1$, then the representational module would adapt; when $\alpha_1\alpha_2<1$ the relational module adapts (see Methods). This prediction is a good fit for the regularized simplified model of Fig. \ref{fig:Lin_a1_to_a2_explain}c, in which $\lambda=0.1$. This prediction also fits the adaptation pattern of the ANN in the more complex images task, when adjusted such that $\alpha_1\alpha_2>\bar\alpha^2$ is the condition for adapting the representational module (Fig. \ref{fig:concept_ada_ratio}b and S5). Overall, our results suggest that the multiplication of $\alpha_1\alpha_2$, which precisely determines the adaptation pathway in the limit of weak regularization, is a good fit to to how $\alpha_1$ and $\alpha_2$ affect the adaptation pathway in this model.

\section*{Discussion}

In this study, we explored how artificial neural networks (ANNs) resolve relational inconsistencies when faced with violations of expected relationships. Our findings reveal two distinct adaptation pathways that naturally emerge from gradient-based learning dynamics: when violations are small, networks primarily adjust their relational expectations, whereas extreme violations lead to modifications in object representations, preserving the initial relational expectation. This dichotomy can account for the experimentally observed inverted U-shaped dependence of expectation adaptation on violation magnitude, where moderate expectation violations lead to learning, but extreme violations often result in resistance to change.

The key component of our theoretical finding is that a violation of expectation can be resolved in more than just one way. Therefore, in case of relational inconsistency, whether or not the relational component will eventually adapt depends on whether adaptation of this component is ``sufficiently fast'', compared with the alternatives route for adaptation -- the adaptations of the representational module. In response to the violation of expectation, a ``race'' between the two adaptation pathways begins. The ratio between the new rule and the original rule determines the starting points of the two adaptation pathways, and hence the distance to the finish line. The larger this ratio, the smaller is the distance required for adaptation by the relation module, relative to the representational module. Moreover, despite the fact that we used the same learning rule for the representational and relation modules, SGD, which is characterized by a single learning rate, the ``speed'' of adaptation of the two modules is not equal. The larger the magnitude of the ``new'' rule ($\alpha_2$) the slower the relation module adapts relative to the representational module. 

Our results indicate that the product of the initial and final rules $\alpha_1\alpha_2$ is a crucial parameter in determining the adaptation pathway. When this product is large, the representational module adapts and when it is small the relational module adapts. In the real world, the rule sizes correspond to an objective difference between objects. Extrapolating our results to the scientist example, if particles A are expected to be larger than particles B by $\alpha_1$, and the novel observation is that they are smaller by $\alpha_2$, then the observation would be dismissed if $\alpha_1\alpha_2$ would be larger than some threshold that depends on the scale $\alpha$.

In our study, the core relationships we investigated were defined by changes in specific predictive features: shapes size in the main manuscript, and grayscale color and number of shapes in the Supplementary Materials. All other features varied randomly between image pairs and were categorized as irrelevant features. Our prior research has demonstrated that the quantity of these irrelevant features directly correlates with the difficulty of similar tasks \cite{barak_naive_2023}. Consequently, we integrated these irrelevant features into our experimental design to slightly increase task complexity, aiming for a more realistic scenario. We hypothesize that as the networks achieve high performance, they learn to effectively disregard these irrelevant features. Crucially, when the rule is reversed in our experiments, only the predictive feature's rule is altered. Therefore, we do not anticipate that the number or identity of these irrelevant features will influence our reported results.

Reconciling prior beliefs with conflicting evidence is commonly framed within a Bayesian framework \cite{knill_bayesian_2004, tenenbaum_how_2011}. To connect our findings to this approach, consider a scenario with two competing hypotheses: one suggesting the predictive feature (e.g., size) increases, and the other suggesting it decreases. When an agent begins with a stronger belief in the "increase" hypothesis and then encounters evidence challenging this belief, Bayesian updating provides a formal way to revise these beliefs. This revision depends on two key factors: the initial confidence in each hypothesis (prior beliefs) and how well each hypothesis explains the new observation (likelihood of the evidence). The Bayesian framework provides a clear decision threshold -- the point at which belief should shift from one hypothesis to the other. For the Bayesian model to account for the inverted U-shaped dependence of relational adaptation to the size of the violation, observed experimentally as well as in our ANN model, we need a model in which large $\alpha_1$ and $\alpha_2$ support belief consistency. If we interpret the magnitudes of $\alpha_1$ and $\alpha_2$ as measures of the strengths of evidence they provide, it is easy to see why a strong initial evidence, in the form of a large $\alpha_1$ would do that. However, it is challenging to interpret belief consistency when $\alpha_2$ is large. This is because stronger contradictory evidence (larger $\alpha_2$) is expected to increase the likelihood of hypothesis revision rather than decrease it. It is possible to account for the inverted U-shaped dependence in a Bayesian framework if we add a measure of confidence in the observations, and posit that the confidence in the first set of observations (associated with $\alpha_1$) is larger than that of the second set of observations (associated with $\alpha_2$).

Traditional psychological perspectives often interpret belief persistence in the face of contradictory evidence as a cognitive bias \cite{lord_biased_1979, kahneman_thinking_2011}. Our findings, however, suggest an alternative view: rather than mere bias, this persistence may reflect an adaptive learning strategy that stabilizes relational expectations by integrating violations through representational adjustments. A more recent theory, \textit{ViolEx} \cite{w_expectancies_2015, pinquart_why_2021, panitz_revised_2021, rief_using_2022}, posits that when violations are extreme, immunization mechanisms act to devalue or reframe the new information. Our findings are consistent with ViolEx, but suggest that the immunization mechanism naturally emerges from the same learning process that drives expectation updates, with the outcome determined by which adaptation pathway ``wins the race.''

In the paper, we found that intermediate adaptation steps can be used to reshape the adaptation pathway. Specifically, a small intermediate step promoted the adaptation of the relational expectation. This result aligns with findings in cognitive-behavioral therapy and education, where gradual exposure to conflicting information enhances adaptive outcomes. For example, therapies designed to modify dysfunctional expectations, such as exposure therapy for anxiety disorders, benefit from structured interventions that introduce moderate violations instead of extreme ones \cite{foa_emotional_1986, barlow_anxiety_2002}. Similarly, in education, conceptual change is more effective when scaffolded gradually rather than introduced through abrupt contradiction \cite{bruner_toward_1974, vygotskij_mind_1979, gopnik_reconstructing_2012}.

The two adaptation pathways discussed in this work are categorical (representational vs. relational). But notably, there are multiple ways to adapt, even within each category. Specifically, the representational module is characterized by a large number of parameters, more parameters than examples. Therefore, there can be multiple combinations of parameters that, for the same set of examples, yield the same representational adaptation. The relational module in our model is rather simple, but in a more general model we expect a similar multiplicity of possible adaptation pathways within the relational pathway. Indeed, different adaptation pathways within a module have been a subject of research in the cognitive sciences. For example, an inconsistency between a person's unhealthy habit of smoking and the warning against the harmful effects of smoking presented on a tobacco package can be resolved in several ways (that do not change the habit): The smoker can posit that benefits associated with appetite suppression outweigh the cancer-related health risks, or alternatively, question the research that links smoking to increased mortality \cite{festinger_theory_1957}. Both solutions are consistent with a change to the representation of smoking. A major limitation of our model is that it is not informative about the determinants of \textit{within-category} adaptation pathways.

Another limitation of our model is that it modeled passive agents who get a sequence of pairs of inputs and are required to learn their relationships. However, humans actively engage in the learning process by performing various comparison and manipulation patterns of stimuli \cite{ribes-inesta_comparison_2020, leon_rbdt_2021}. This active participation is useful for establishing relational behavior. The active manipulation of stimuli can be interpreted as a form of representational adaptation, where the learner actively re-frames or re-processes the sensory input by physically interacting with it, allowing the learner to adjust their internal representations of objects to resolve inconsistencies and make new information compatible with existing or emerging relational rules.

By examining the intrinsic learning dynamics of neural systems, our work provides insights into the mechanisms that govern adaptation to inconsistencies. The competition between representational and relational adaptation pathways naturally produces the non-monotonic patterns of belief updating observed in human cognition. These findings have implications for understanding learning processes across cognitive, educational, and therapeutic contexts. Further work is needed to explore the nuances of adaptation within the representational and relational categories, as well as to experimentally test the model's predictions.

\section*{Methods}

\subsection*{Code availability}

A PyTorch \cite{paszke_pytorch_2019} code that generates the results and figures of this paper is available at:\newline \url{https://github.com/Tomer-Barak/relational_expectation_violations}.

\subsection*{Order discrimination task}

The order discrimination task was designed to assess the ability of ANNs to determine the correct order of image pairs based on a specific feature. As written in the main text, each image in the pair depicted shapes arranged on a $3\times 3$ grid and was characterized by five features: grayscale color, number of shapes, size, grid arrangement, and shape type. The images were $224\times 224$ pixels in size and grayscale (they consisted of 3 channels with identical values). The size of the shapes was defined as the diameter of the circle enclosing them.

The "correct" order in the task was determined by the identity of the relevant feature (color, size, or number), termed the \textit{predictive feature}, and whether this feature increased or decreased from left to right. To construct the training set, the predictive feature values were randomly selected from a uniform distribution over possible values for the left image. The corresponding values for the right image were then calculated by applying the rule parameter $\alpha$ to the left image's values. Non-predictive features were randomly selected from a uniform distribution for each image pair, remaining constant within the pair.

For each $\alpha$ and a tested network, we constructed a training set that consisted of 160 image pairs that demonstrated that rule. To evaluate the performance of an ANN in this task, we tested its ability to classify the correct order of $32$ novel image pairs. In Figs. \ref{fig:task_eval}a-b and \ref{fig:task_switch}b, we averaged the classification accuracy of 100 ANNs and estimated the confidence interval based on the standard error.

In the main text of this paper, we presented the results when the predictive feature is the size. Similar results were obtained when the predictive features were color or number, and these are presented in the Supplementary Information (Figs. S1-S4). For full implementation of the task, see Images.py in the paper's GitHub site:\newline \url{https://github.com/Tomer-Barak/relational_expectation_violations}

\subsection*{The ANN}

The representational module $Z_\*{w}(\mathbf{x})$ consisted of three convolutional layers (number of filters: 16, 32, 32; kernel sizes: 2, 2, 3; strides: all 1; padding: all 1) followed by one fully-connected linear layer (taking a $2592$-dimensional vector to a one-dimensional output). Three ReLU activation functions were applied after each convolutional layer, and two Max-Pool layers (kernels: 4 and 6, strides: all 1) were applied after the second and third convolutional (+ReLU) layers. The parameters of $Z_\*{w}$ were randomly initialized using PyTorch's \cite{paszke_pytorch_2019} default initialization (uniform distribution scaled by $1/\sqrt{N}$ where $N$ is the number of the layer's input neurons).

Given a training set, we optimized the randomly initialized ANN's parameters to minimize the regularized loss function \eqref{eq:loss_reg} with the vanilla SGD optimizer ($lr=0.004$). For the regularization term, we used the hyperparameters $\lambda=4$ and $r^2=0.1$. We used a batch size of $2$ image pairs and applied $20$ optimization steps per batch.

To assess the adaptation pathway of an ANN, we measured its parameter $\theta$ during training. To complement this measure, we also measured the average $\Delta Z$ of the ANN over $32$ test image pairs from the same training set distribution (with the same $\alpha$). A network that changed its sign of $\theta$ and kept the sign of $\Delta Z$ after rule reversal was classified as adapting its relational module. A network that kept the sign of $\theta$ while changing the sign of $\Delta Z$ has adapted its representational module. We excluded networks that kept or changed the signs of both $\theta$ and $\Delta Z$ together. These networks necessarily failed the task. The fraction of excluded networks was less than $1\%$: Fig. \ref{fig:concept_ada_ratio}a: $3/1800$. The fraction of networks that adapted their relational module (e.g., in Fig. \ref{fig:concept_ada_ratio}a) was obtained by $ \#\theta/\left(\#Z + \#\theta\right)$ where $\#Z$ is the number of networks that adapted $\Delta Z$, and $\#\theta$ is the number of networks that adapted $\theta$.

\subsection*{Calculating the inflection point in Figure \ref{fig:concept_ada_ratio}a}

To calculate the inflection point $\bar\alpha$, we fitted a logistic function to the results of how many networks adapted their relational module as a function of $\alpha$. Specifically, we fitted the two parameters $c$ and $d$ of the logistic function $\frac{1}{1 + e^{c\left(\alpha - d\right)}}$. The inflection point was defined as $\bar\alpha=d$. The $95\%$ CIs of $\bar\alpha$ correspond to $1.96\cdot SE(d)$ where $SE(d)$ is the standard deviation error of the estimation of $d$ using SciPy's \cite{virtanen_scipy_2020} curve fitting function.

\subsection*{Simplified model: two attractive fixed points}
Because this is a gradient system, the dynamics will necessarily converge to the (stable) fixed point(s) of the dynamics. To find the fixed point(s), we consider the two nullclines, $\dot{\Delta Z}=0$ and $\dot\theta=0$. From these equations we write,
\begin{equation}
\begin{split}
\left|\left( \Delta Z  - \theta \right) \right|&= 2\lambda \left|\left( \Delta Z ^2 + \theta^2 -r^2 \right)\right|\left|\Delta Z\right|  \\
\left| \left( \Delta Z - \theta \right) \right|&= 2\lambda \left|\left( \Delta Z  ^2 + \theta^2 -r^2 \right)\right|\left|\theta\right|.
\end{split}
\end{equation}
Subtracting the equations, we get that $\left|\left( \Delta Z  ^2 + \theta^2 -r^2 \right)\right|\left(\left|\Delta Z\right|-\left|\theta\right|\right)=0$. If $\left|\left( \Delta Z  ^2 + \theta^2 -r^2 \right)\right|=0$ then from Eq. \eqref{eq:dynamics}, $\Delta Z=\theta$ at the fixed point. Therefore together, $\left|\Delta Z\right|=\left|\theta\right|$.

The nullclines are depicted in Fig. \ref{fig:quiver}. The fixed points can be computed analytically by substituting $\left|\Delta Z\right|=\left|\theta\right|$ in the nullclines equations. We find that there is a trivial fixed point at $\Delta Z=r=0$. A linear stability analysis reveals that this fixed point is unstable. Additionally, there are two additional fixed points $\Delta Z=\theta=\pm \frac{r}{\sqrt 2}$. These fixed points satisfy both $\mathcal{L}$ and the regularization term. We will discuss their stability shortly. When the regularization term is large, $\lambda>\frac{1}{r^2}$, there are two additional fixed points, $\Delta Z=-\theta=\pm \sqrt\frac{r^2-\frac{1}{\lambda}}{2}$, but a linear stability analysis reveals that they are unstable. Because the dynamics is driven by a gradient of a loss function, then it necessarily converges to a fixed point. Because the $\Delta Z=\theta=\pm \frac{r}{\sqrt 2}$ are the only non-unstable fixed points, they are necessarily the only attractors of the dynamics.

\subsection*{Weakly regularized simplified model: Exact solution}

To understand how the magnitude of $\alpha$ affects this adaptation pathway, it is useful to consider the dynamics of a weakly regularized system, where $\lambda\ll1$. In this case, the dynamics first minimize the unregularized part of the loss, $\left(\Delta Z - \theta \right)^2$, driving the system to $\Delta Z = \theta$, and then the regularization kicks in to set the system on the ring $\Delta Z^2=\theta^2=r^2/2$.

Without regularization, the dynamical equations simplify to
\begin{equation}
\begin{split}
    \dot{\Delta Z} &=  -\alpha^2 \left( \Delta Z -  \theta \right)  \\
    \dot{\theta} &=  \left( \Delta Z - \theta \right).
\end{split}
\end{equation} 
These two equations are linearly dependent, implying that the unregularized system converges to a point on a line attractor that depends on its initial state. Specifically, due to the equations being linearly dependent, the value $\Delta Z+\alpha^2\theta$ is conserved during optimization and its value depends on the initial state $\Delta Z(0) + \alpha^2\theta(0)$. This is true also at the fixed points, where $\Delta Z^* = \theta^*$. Plugging the fixed point solution to the conservation law provides the exact point the system would reach on the line attractor $\Delta Z = \theta$:
\begin{equation}
   \Delta Z = \theta =  \frac{\alpha^2 \theta (0)+ \Delta Z (0)}{\alpha^2+1},
\end{equation}

In the rule reversal case, assuming that the system starts from the positive fixed point, the initial values are $\Delta Z (0) = - r/\sqrt{2} $ and $\theta = r/\sqrt{2}$. Substituting this initial state, we find that the unregularized system is driven to the following point on the line attractor
\begin{equation}
\label{eq:dynamics_linear_reverse_fp}
   \Delta Z^* = \theta^* =  \frac{r}{\sqrt{2}}\frac{\alpha^2-1}{\alpha^2+1}.
\end{equation}

When approaching the the line attractor, $\left(\Delta Z - \theta \right)^2$ becomes small, comparable to the regularization term in Eq. \eqref{eq:dynamics}. Therefore, the regularization would then become more dominant and drive the system towards $\Delta Z^{*2}=\theta^2=r^2/2$. The result we arrived at, Eq. \eqref{eq:dynamics_linear_reverse_fp}, shows that whether $\alpha^2$ is smaller or larger than $1$ determines the sign of the fixed point. $\alpha^2>1$ corresponds to a fixed point where both $\Delta Z$ and $\theta$ are positive, keeping the original sign of $\theta$, whereas $\alpha^2<1$ leads to a negative fixed point, changing the sign of $\theta$. Therefore, the value of $\alpha^2$ distinguished between the two adaptation pathways, and the inflection point is at $\bar\alpha=1$. We verified this analysis by simulating the simplified model with weak regularization. For example, Fig. \ref{fig:quiver} demonstrates the dynamics when $\lambda=0.1$ for a strong violation $\alpha>\bar\alpha$ and a weak violation $\alpha<\bar\alpha$. Initially, the dynamics drive the system to the line $\Delta Z = \theta$, minimizing the unregularized term by either changing the sign of $\Delta Z$ or $\theta$, depending on the size of the violation. Then, when $\Delta Z \approx \theta$, the regularization term pushes the system towards one of the two fixed points, where $\Delta Z^{*2}=\theta^2=r^2/2$.

In the more general case, where the rule changes from $\alpha_1$ to $-\alpha_2$, the initial state of $\Delta Z$ before the adaption changes. To see this, remember that $\Delta Z=w\alpha$ where $\alpha$ is the current rule. At the first learning phase, assuming that the system converged to the positive fixed point, the value of the representational module's weight at the fixed point, $w^*$, is given by $\Delta Z^* = w^* \alpha_1 = r/\sqrt 2$. When flipping the rule, $w^*$ remains as it is, while $\Delta Z$ is now defined with $\alpha_2$. Therefore, $\Delta Z (0) = -w^*\alpha_2 = -\frac{r}{\sqrt 2}\frac{\alpha_2}{\alpha_1}$. The point on the line $\Delta Z = \theta$ where the system approaches depends on this initial state (Eq. \eqref{eq:dynamics_linear_general_fp}):
\begin{equation}
\label{eq:dynamics_linear_a1_a2_fp}
   \Delta Z^* = \theta^* =  \frac{r}{\sqrt{2}}\frac{\alpha_2^2-\frac{\alpha_2}{\alpha_1}}{\alpha_2^2+1}.
\end{equation}
This equation shows that whenever $\alpha_1\alpha_2>1$, the system adapts its representational module, while for $\alpha_1\alpha_2<1$ it would adapt its relational module. We verified this prediction in the weakly regularized simplified model in Figures \ref{fig:concept_ada_ratio}b and S5.

\section*{Data availability statement}

The data that were used for the order discrimination tasks was generated in real-time by an algorithm. The generating code is available on this project's GitHub page:\newline
\url{https://github.com/Tomer-Barak/relational_expectation_violations}.

\bibliography{citations}

\section*{Acknowledgements}

This work was supported by the Gatsby Charitable Foundation.
Y.L. holds the David and Inez Myres Chair in Neural Computation.
We thank David Hansel and Ilya Nemenman for insightful discussions. 

\section*{Author contributions statement}

T.B. conducted the experiments, T.B. and Y.L. designed the study, analyzed the results and wrote the manuscript.

\section*{Additional information}
\subsection*{Competing interests statement}

The authors declare no competing interests.

\end{document}


\maketitle

\thispagestyle{empty}

\section*{The results for other predictive features: Color and Number}

\begin{figure}[H]
\vskip -0.2in
\begin{center}
\centerline{\includegraphics[width=0.9\linewidth]{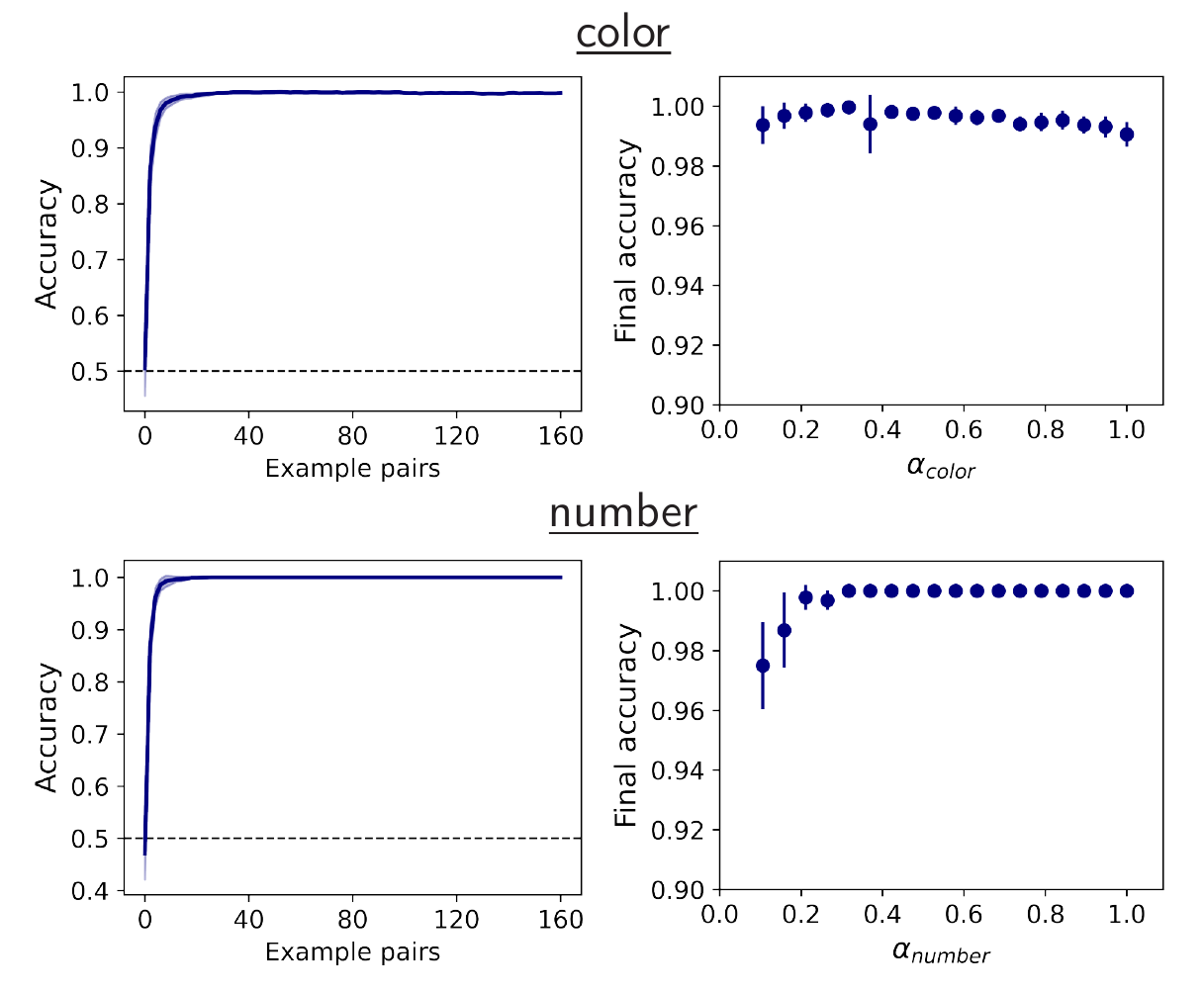}}
\caption{\textbf{Task performance.} Left: The average test accuracy of 100 networks trained on a task where the predictive features were color (top) or number (bottom). The change rule was $\alpha=0.5$. Right: The final test accuracies for various values of $\alpha$, averaged over 100 networks per $\alpha$. Error shades and bars correspond to $95\%$ CI.}
\label{fig:Just_training_other_PFs}
\end{center}
\vskip -0.3in
\end{figure}

\begin{figure}[H]
\begin{center}
\centerline{\includegraphics[width=\linewidth]{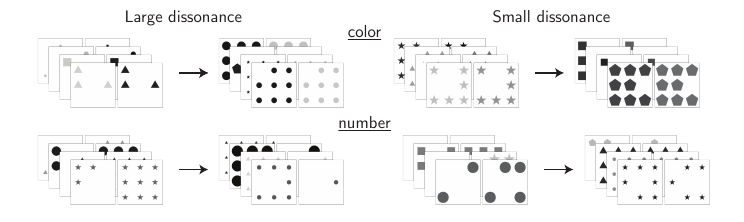}}
\caption{\textbf{Simulating a dissonance.} Initially, the predictive feature (top: color, bottom: number) increases by $\alpha$. Then, the relationship is reversed to $-\alpha$. The dissonance magnitude is represented by $\alpha$. Left: large dissonance $\alpha=0.8$; Right: small dissonance $\alpha=0.2$.}
\label{fig:task_switch_other_PFs}
\end{center}
\end{figure}

\begin{figure}[H]
\begin{center}
\centerline{\includegraphics[width=\linewidth]{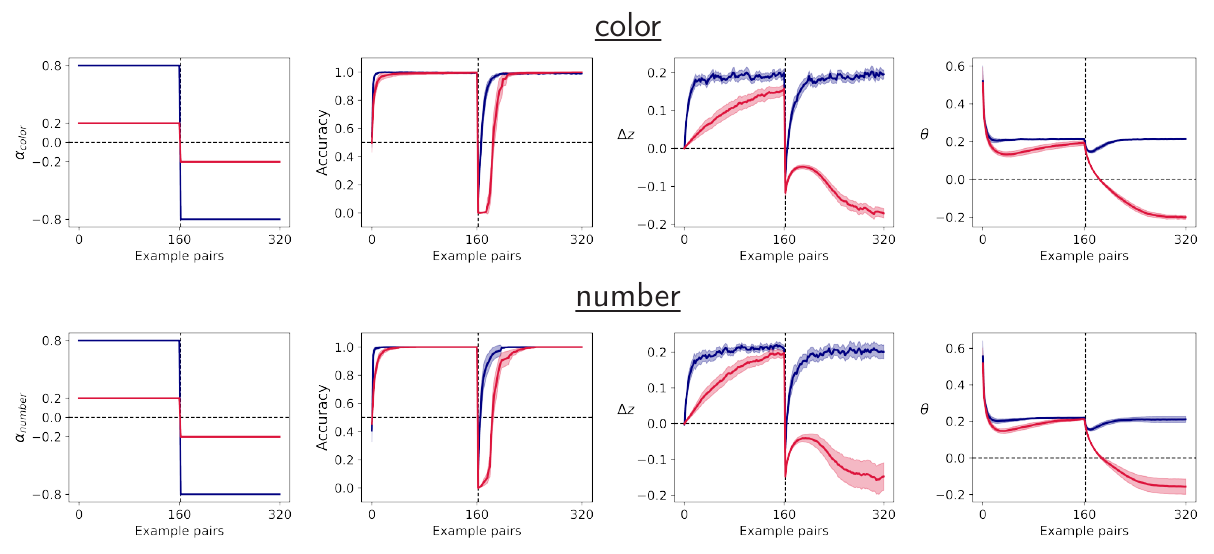}}
\caption{
\textbf{Performance and adaptation following rule reversal.} 100 networks were presented with $160$ examples in which $\alpha=0.8$ (blue) or $\alpha=0.2$ (red). Then, we reversed this rule, simulating a cognitive dissonance. From left to right, as a function of training examples: The rule $\alpha$; Classification accuracy of the networks during the task; The ANNs' $\Delta Z$; The ANNs' $\theta$. The lines are the averages, and the shades correspond to $95\%$ CI.}
\label{fig:two_alphas_other_PFs}
\end{center}
\end{figure}

\begin{figure}[H]
\begin{center}
\centerline{\includegraphics[width=0.6\linewidth]{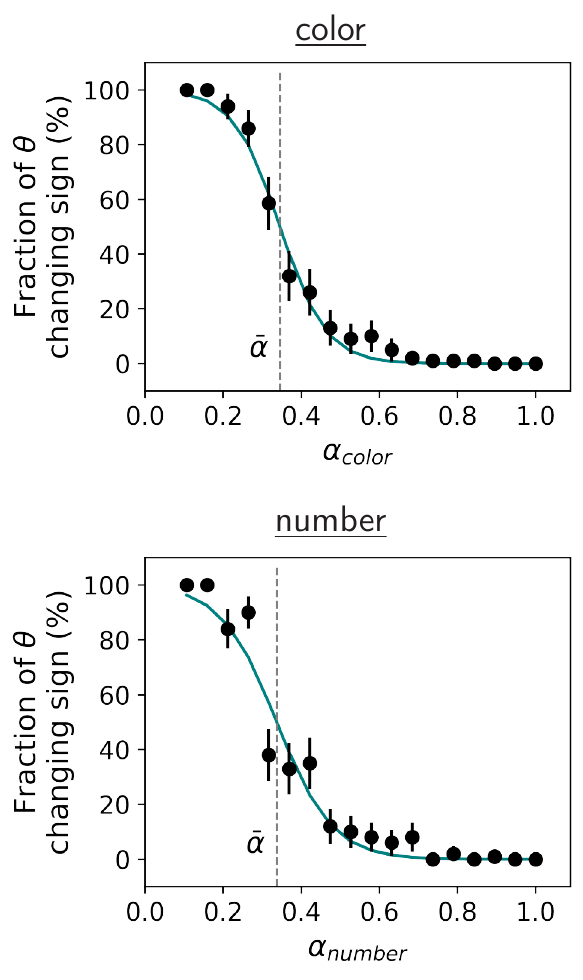}}
\caption{\textbf{Adaptation pathway versus dissonance magnitude.} For the color (left) and number (right) predictive feature, the percentage of networks that adapted their input representation $\Delta Z$ to match the expected $\theta$ increases a function of $\alpha$. The inflection points between the adaptation pathways, $\bar\alpha$, were obtained by fitting the results (black) with a logistic function (green). Error bars correspond to $95\%$ CI (Wilson estimation).}
\label{fig:Concept_ratios_other_PFs}
\end{center}
\end{figure}

\section*{Fitting the adaptation pattern of the ANN}

\begin{figure}[H]
\begin{center}
\centerline{\includegraphics[width=0.6\linewidth]{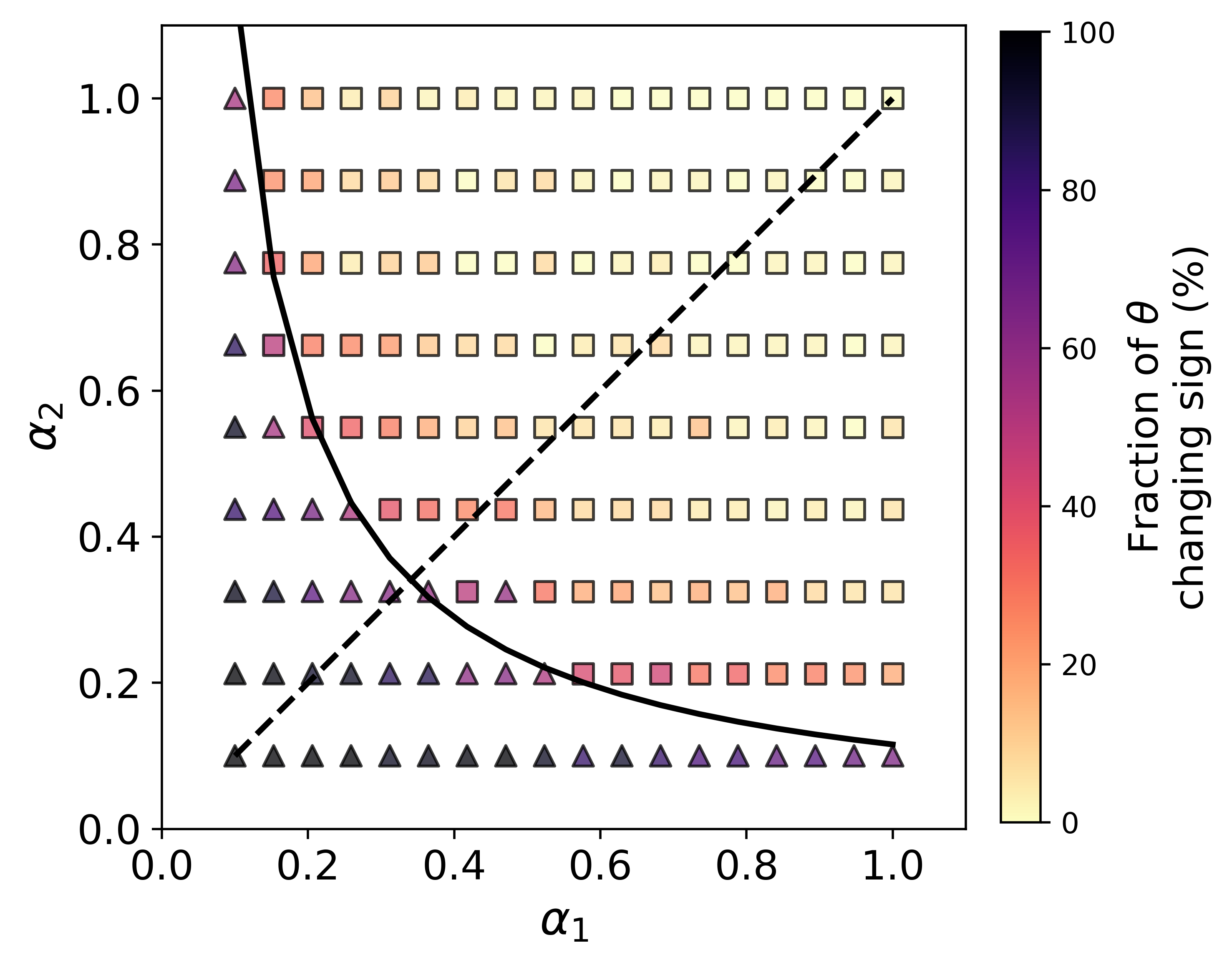}}
\caption{\textbf{Adaptation patterns for general rule reversals for the ANN.} The adaptation pattern, taken from (Fig. 5), with the predicted inflection line $\alpha_1\alpha_2=\bar\alpha^2=0.34^2$ (black solid line). The value $\bar\alpha$ was obtained from the symmetric rule reversal case, represented by the dashed black line.}
\label{fig:ANN_a1_to_a2_fit}
\end{center}
\end{figure}